\documentstyle[12pt,fleqn]{article}
\catcode`\@=11
\@addtoreset{equation}{section}

\def\be{\begin{equation}}
\def\ee{\end{equation}}
\def\ba{\begin{eqnarray}}
\def\ea{\end{eqnarray}}

\def\rbar{\overline{r}}
\def\rhbar{\overline{\rho}}

\def\la{\mathrel{\mathpalette\fun <}}

\def\fun#1#2{\lower3.6pt\vbox{\baselineskip0pt\lineskip.9pt
        \ialign{$\mathsurround=0pt#1\hfill##\hfil$\crcr#2\crcr\sim\crcr}}}

\begin{document}
\begin{titlepage}
\null\vspace{-62pt}
\begin{flushright}FERMILAB-Pub-93/378-A\\
January, 1994\\
(revised version)
\end{flushright}

\vspace{.2in}
\baselineskip 24pt
\centerline{\large \bf{Mass and radius of cosmic balloons}}

\vspace{.5in}
\centerline{Yun Wang}
\vspace{.2in}
\centerline{{\it NASA/Fermilab Astrophysics Center}}
\centerline{\it Fermi National Accelerator Laboratory, Batavia, IL 60510-0500}

\vspace{.7in}
\centerline{\bf Abstract}
\begin{quotation}

Cosmic balloons are spherical domain walls with relativistic particles trapped
inside. We derive the exact mass and radius relations for a static cosmic
balloon using Gauss-Codazzi equations.
The cosmic balloon mass as a function of its radius, $M(R)$, is found to have a
functional form similar to that of fermion soliton stars, with a fixed point at
$2GM(R)/R\simeq 0.486$, which corresponds to the limit of infinite central
density.
We derive a simple analytical approximation for the mass density of a
spherically symmetric relativistic gas star. When applied to
the computation of the mass and radius of a cosmic balloon, the analytical
approximation yields fairly good agreement with the exact numerical solutions.

\end{quotation}

\end{titlepage}

\baselineskip=24pt

\section{1. Introduction}

A particle with a discrete symmetry may have a different mass on either side of
the domain wall, which may form in an early Universe phase transition with the
spontaneous breaking of the discrete symmetry. When a particle has less energy
than the rest mass it would have on the other side of the wall, it can become
trapped inside a region with a closed domain wall as the boundary. Closed
domain walls filled with such particles have been named ``cosmic balloons''
[\ref{CB}]. Due to effects such as the emission of gravitational radiation,
irregular shapes of balloons oscillate and eventually settle down to spheres.
Here we are only concerned with a spherical cosmic balloon.

Cosmic balloons are stable objects. The pressure of the relativistic gas inside
the balloon balances the surface tension of the wall. An incarnation
of a cosmic balloon is a neutrino ball [\ref{NB}], which is a spherical domain
wall with light right-handed neutrinos trapped inside. Details of neutrino
balls have been worked out which seem to indicate that they are plausible
astrophysical objects [\ref{NB}], possibly providing an alternative
explaination for the mass of quasars and other astrophysical phenomena.

In this work, we follow Ref[\ref{CB}] in the study of general properties of
cosmic balloons, in particular, the mass of the balloon as a function of its
radius. The analytical approximation in
Ref[\ref{CB}] has an extremely involved numerical form, and the mass versus
radius curve found there is numerically incomplete. In another previous work
with similar objectives, Ref[\ref{Po}], an analytical approximation was applied
beyond its valid range and led to incorrect numerical solutions of the mass and
radius. In this work, we find simple and transparent analytical expressions
which are useful in helping us understand
the cosmic balloon solutions qualitatively, and
we compute the complete set of mass and radius numerically with and without
the application of the analytical approximation.

\section{2. Mass and radius relations}

To find the radius $R$ and mass $M$ of a cosmic balloon, we study the static
configuration of a spherical domain wall containing gas. Let the metric inside
the domain wall ($r<R$) be
\be
d \tau^2= B(r) dt^2- A(r) \, dr^2 -r^2 (d\theta^2+ \sin^2\theta d\phi^2).
\ee
The motion of spherical domain walls containing vacuum has been solved in
Ref.(\ref{SIKIVIE}), using the Gauss-Codazzi
formalism. Following Ref.(\ref{SIKIVIE}), we find the
equations of motion for a spherical domain wall containing a perfect fluid
with pressure $p$ and density $\rho$:
\ba
(\alpha+\beta)\ddot{R}+\frac{\alpha G M}{R^2}+ \frac{\beta}{2} \left[
\frac{A'}{A} \,\dot{R}^2+\frac{B'}{B}\, \alpha^2\right] &=&
-\alpha\beta \left[\frac{2(\alpha+\beta)}{R} -\frac{2p}{\sigma}
-\frac{2\dot{R}^2}{\sigma}\, A(\rho+p)\right] \nonumber\\
(\alpha-\beta)\ddot{R}+\frac{\alpha G M}{R^2}- \frac{\beta}{2} \left[
\frac{A'}{A} \,\dot{R}^2+\frac{B'}{B}\, \alpha^2\right] &=&
-4\alpha\beta\pi G \sigma,
\label{eq:static}
\ea
Here a dot denotes a derivative with respect to proper time, and a prime
denotes a derivative with respect to $r=R$.
We have defined
\ba
&& \alpha \equiv \left[A^{-1}(R)+\dot{R}^2\right]^{1/2}, \nonumber\\
&& \beta \equiv \left[ 1-\frac{2GM}{R} +\dot{R}^2 \right]^{1/2}.
\ea
$M$ is the total mass of a cosmic balloon of radius $R$.

The equation for hydrostatic equilibrium is [\ref{SW}]
\be
\frac{B'}{B}=-\frac{2p'}{p+\rho}.
\ee
This and the Einstein equations can be combined to give a single differential
equation:
\be
\label{eq:basic}
-r^2 \, p'(r)= G {\cal M}(r) \rho(r)
\left[ 1+\frac{p(r)}{\rho(r)} \right] \left[ 1+\frac{4\pi r^3 p(r)}
{{\cal M} (r)} \right] \left[1- \frac{2G {\cal M}(r)} {r} \right]^{-1},
\ee
where we have defined ${\cal M}(r)$ such that
\be
{\cal M}'(r) = 4 \pi r^2 \rho(r), \hskip 2cm  {\cal M}(0)=0.
\ee
For $A(0)$ finite,
\be
A(r)= \left[ 1- \frac{2G {\cal M}(r)}{r} \right]^{-1}.
\ee
Using the above well known results in Eqs.(\ref{eq:static}) and setting
$\dot{R}=0=\ddot{R}$, we find
\ba
\label{eq:M(R)1}
&&M(R)= {\cal M}(R)+ 2\pi R^2 \sigma (\alpha+\beta), \\
&& p(R)= \frac{\sigma}{2R} \, \left(3\alpha+
\frac{1}{\beta}\right),
\label{eq:R}
\ea
with
\be
\alpha=\left[1-\frac{2G{\cal M}(R)}{R}\right]^{1/2},
\hskip 2cm \beta=\left[1-\frac{2G M}{R}\right]^{1/2}.
\ee
Eq.(\ref{eq:M(R)1}) can be rewritten to give
\be
\beta = \alpha-4\pi R G \sigma, \hskip 2cm {\rm for}\,\,\,
R<\frac{\alpha}{4\pi G \sigma}.
\ee
A cosmic balloon with $R>\alpha/(4\pi G \sigma) $ is contained within its
Schwarzschild radius, a case in which we are not interested here.
Hence
\be
\label{eq:M(R)}
M(R)= {\cal M}(R) +4\pi R^2 \sigma [\alpha-2\pi R G \sigma].
\ee
Note that ${\cal M}(R)$ is the total energy of the matter $and$ the
gravitational field inside the cosmic balloon [\ref{SW}].
Eq.(\ref{eq:R}) gives the radius $R$ of a cosmic balloon as a function
of its surface tension $\sigma$, it can be rewritten as
\be
p(R)=\frac{2 \alpha \sigma}{R} +\frac{G\sigma {\cal M}(R)}{\beta R^2}
+ \frac{\alpha}{\beta} \, 2\pi G \sigma^2.
\ee
We see that the gas pressure at $R$ is balanced by the surface tension
of the domain wall, the gravitational force on the wall due to the total
mass inside the wall and the wall itself.

\section{3. An exact solution}

Let us define dimensionless variables as follows:
\ba
\label{eq:bars}
x(r) &\equiv& \frac{2G{\cal M}(r)}{r}, \nonumber\\
\rbar &\equiv& \frac{r}{r_0}, \hskip 2cm
\left(r_0 \equiv [48\pi G\sigma]^{-1}\right),\nonumber\\
\rho^{*}(r) &\equiv& \frac{\rho(r)}{\rho_0}, \hskip 2cm
\left(\rho_0 \equiv [8\pi G r_0^2]^{-1}\right), \nonumber\\
\overline{M}(R) &\equiv& \frac{M(R)}{M_0}, \hskip 2cm
\left(M_0 \equiv \frac{r_0}{2G}\right).
\ea
Note that the length scale $r_0$ and the mass scale $M_0$ can be written as
\ba
&& r_0 \simeq 1.95 \times 10^{13} {\rm cm} \left[ \frac{\sigma}
{({\rm TeV})^3} \right]^{-1}, \nonumber \\
&& M_0 \simeq 6.6 \times 10^{7} M_{\odot} \left[ \frac{\sigma}
{({\rm TeV})^3} \right]^{-1}.
\ea
This is interesting since both $\overline{R}$ and $\overline{M}(R)$ have
maxima on the order of $1$, as we shall see.
$\sigma^{1/3}$ is usually
associated with the scale of the discrete symmetry breaking, but it could be
made anything depending on the particular model of cosmic balloon.

Eq.(\ref{eq:basic}) can be rewritten as
\be
\label{eq:main}
 x''= - \frac{2}{3} \, \frac{1}{1-x} \,
\left[ x'^2 +5 x' \, f_1(\rbar)
+(7x-3) f_2(\rbar) \right],
\ee
where the primes denote differentiation with respect to $\rbar$, and
\ba
&&f_1(\rbar) \equiv \frac{x}{\rbar}, \hskip 2cm f_1(0)=0,\nonumber\\
&&f_2(\rbar) \equiv \frac{x}{\rbar^2}, \hskip 2cm f_2(0) \equiv q=
\frac{8\pi G \rho(0) r_0^2}{3}, \nonumber \\
&&x(0)=0= x'(\rbar=0),  \hskip 2cm {\rm if}\,\, \rho(0) \,\,
{\rm is \,\,finite}.
\ea
Note that $\rho(0)$ dependence only comes in through the boundary condition of
the function $f_2(\rbar)$. In terms of $x$ and $\rbar$,
\be
\rho^*(r)=\frac{1}{\rbar^2} \left[x+\rbar \,x' \right].
\ee
The scaled radius of the cosmic balloon $\overline{R}$ is given by the
intersection of $\rho^*(r)$ and
\be
\rho^*(R)=\frac{1}{4\overline{R}} \left(3\alpha+\frac{1}{\beta}
\right).
\ee
The scaled mass of the cosmic balloon is
\be
\label{eq:M}
\overline{M}(R) = \overline{R} \left[ x(\overline{R})+
\frac{\overline{R}}{6} \left(\alpha-\frac{\overline{R}}{24}\right) \right].
\ee
Here
\be
\alpha=\left[1-x(R)\right]^{1/2}, \hskip 1.5cm
\beta=\alpha-\overline{R}/12, \hskip 2cm {\rm for}\,\,\,
\overline{R} <12 \alpha.
\ee

Eq.(\ref{eq:main}) has an obvious solution [\ref{SW}]
\[
x =\frac{3}{7}, \hskip 2cm \rho(r) = \frac{3}{56\pi Gr^2} \,\,\,
{\rm or} \,\,\, \rho^*(r)=\frac{3}{7\,\rbar^2},
\]
which has infinite central mass density. The corresponding cosmic balloon
radius and mass are:
\ba
&&\overline{R} = \frac{4}{\sqrt{7}}\left(5-\sqrt{22}\right) \simeq 0.468,
\nonumber\\
&&\overline{M}(R)=\frac{8(122-25\sqrt{22})}{63\sqrt{7}}\simeq 0.2275,
\nonumber\\
&&\frac{2GM(R)}{R}  = \frac{\overline{M}(R)}{\overline{R}}=
\frac{2(20-\sqrt{22})}{63}\simeq 0.486.
\ea

\section{4. Numerical solutions }

The numerical solutions of the cosmic balloon mass function can be obtained in
a straightforward way. It is most convenient to use the parameter $q$
to track the solutions of Eq.(\ref{eq:main}). Note that
\be
\label{eq:q}
q \equiv \frac{8\pi G \rho(0) r_0^2}{3} \propto \rho(0), \hskip 2cm
\rho^*(0)= 3 q.
\ee
For a given value of $q$, Eq.(\ref{eq:main}) can be integrated to find
$x(r)$ and hence $\rho^*(r)$. $\rho^*(r)$ and
$\rho^*(R)=(3\alpha+\beta^{-1})/4\overline{R}$ have
{\it two} intersections for $q >q_c$, which give two values of the cosmic
balloon radius $R$, corresponding to two different branches of the
cosmic balloon mass/radius solutions (see Fig. 1).
The two branches of solutions meet at $q=q_c \simeq 1$, for $q <q_c$
(or $\rho(0) < \rho_c(0)$)
there are no solutions and cosmic balloons do not exist. $q_c$ corresponds
to the minimum of the central mass density for a cosmic balloon.

Fig.2 and Fig.3 are the radius $R$ and mass $M(R)$ versus $q$ respectively.
For increasing $q$ (i.e., $\rho(0)$), the first branch of solutions gives
decreasing $R$ and $M(R)$;
the second branch of solutions gives $R$ and $M(R)$ with complex and similar
behavior. In the second branch of solutions, $R$ increases
until it reaches the maximum ($\overline{R}_m \simeq 0.732$),
then it decreases until it reaches the minimum,
it then increases again until it reaches a second (smaller) maximum, then it
decreases again to a second (larger) minimum, and continues in this manner.
{}From Fig. 2, it is clear that $R$ oscillates with damping amplitude around
$\overline{R}\simeq 0.468$ for large
$q$. $M(R)$ has the same behavior but with a phase lag in $q[\rho(0)]$
compared to $R$ ($\overline{M}_m \simeq 0.374$, $\overline{R}_{Mm}
\simeq 0.71$), and it oscillates around $\overline{M}(R)\simeq 0.2275$ for
large $q$ (see Fig. 3). This is not surprising, since $\overline{R}
\simeq 0.468$ and $\overline{M}(R)\simeq 0.2275$ correspond to the exact
solution of $\rho(r)$ in the limit of
$\rho(0) \rightarrow \infty$ (i.e., $q \rightarrow \infty$). In the plot of
$M(R)$ versus $R$ (Fig. 4), we see that the mass function spirals in toward the
fixed point ($\overline{R}\simeq 0.468$, $\overline{M}(R)\simeq 0.2275$), which
satisfies $2GM(R)/R \simeq 0.486$.

The maximum mass and the corresponding radius for a cosmic balloon are
\ba
&& M_{max} =\overline{M}_m \, M_0 \simeq 2.47 \times 10^{7} M_{\odot} \left[
\frac{\sigma}
{({\rm TeV})^3} \right]^{-1},
\nonumber \\
&& R_{Mmax}= \overline{R}_{Mm} \, r_0 \simeq
1.38 \times 10^{13} {\rm cm} \left[ \frac{\sigma}
{({\rm TeV})^3} \right]^{-1},
\ea
Since $2GM(R)/R=\overline{M}(\overline{R})/\overline{R}$, the fractional red
shift $z$ of a spectral line emitted from the surface of a cosmic balloon
[\ref{SW}]
\be
z \equiv \frac{\Delta \lambda}{\lambda} =\left(1-\frac{2GM}{R} \right)^{-1/2}-1
\ee
is $independent$ of the actual physical scales $r_0$ and $M_0$, i.e.,
independent of  $\sigma$ (the surface tension of the domain wall).
The exact numerical solution shows that $2GM/R$ reaches maximum
($(2GM/R)_{max} \simeq 0.56$) after the mass
has already reached maximum. Before $M=M_{max}$, $2GM/R$ increases
with increasing $M$.
The maximum of mass then corresponds to the maximum of the fractional red shift
$z$ for a stable cosmic balloon, $z_{max} \simeq 0.46$ ($2GM_m/R_{Mm} \simeq
0.53$). At the minimum of the central mass density
($\overline{R}_c \simeq 0.56$, $\overline{M}_c\simeq 0.17$,
$2GM_c/R_c \simeq 0.30$), $z_c \simeq 0.20$.

\section{5. Analytical approximations}

To obtain useful analytical approximations, we scale Eq.(\ref{eq:main}) such
that its solutions are independent of the central mass density $\rho(0)$
(i.e., independent of $q$). Define
\[
t=\sqrt{3q} \, \rbar.
\]
Eq.(\ref{eq:main}) becomes
\be
\label{eq:smain}
\ddot{x}=-\frac{2}{3}\frac{1}{1-x} \left[\dot{x}^2+5\dot{x}f_1^*(t)+
(7x-3)f_2^*(t)
\right],
\ee
where the dots denote differentiation with respect to $t$, and
\ba
&&f_1^*(t) \equiv \frac{x}{t}, \hskip 2cm f_1^*(0)=0,\nonumber\\
&&f_2^*(t) \equiv \frac{x}{t^2}, \hskip 2cm f_2^*(0) \equiv \frac{1}{3},
\nonumber \\
&&x(t=0)=0= \dot{x}(t=0).
\ea
Note that Eq.(\ref{eq:smain}) is exactly the same in form as
Eq.(\ref{eq:main}),
but {\it without} $\rho(0)$ dependence in the boundary conditions.
Correspondingly,
\be
\label{eq:srho}
\overline{\rho}(t) \equiv \frac{\rho^*(r)}{\rho^*(0)}=
\frac{\rho(r)}{\rho(0)}=\frac{1}{t^2}\left[x+t\, \dot{x}\right].
\ee
Combining Eqs.(\ref{eq:srho}) and (\ref{eq:smain}), we can express $x(t)$
in terms of $\overline{\rho}(t)$:
\be
x=\frac{\left[ \left(\dot{\overline{\rho}}/{\overline{\rho}}\right)+
{2}\overline{\rho} \,t /3 \right]
t}{\left(\dot{\overline{\rho}}/{\overline{\rho}}\right) t-2}.
\ee
The radius of the cosmic balloon is now given by
\be
\label{eq:Rq0}
\rhbar(t_R)=\frac{1}{4\sqrt{3q} \, t_R}\left(3\alpha(t_R)+\frac{1}{\beta(t_R)}
\right),
\ee
where $t_R \equiv \sqrt{3q} \, \overline{R}$, and
\ba
&&\alpha(t_R)=\left[1-x(t_R)\right]^{1/2}, \nonumber\\
&&\beta(t_R)=\alpha(t_R)-\frac{t_R}{12\sqrt{3q}}.
\ea
Given the function $\rhbar(t)$, we can find $R(q)$.
The mass of a cosmic balloon is (see Eq.(\ref{eq:M}))
\be
\label{eq:Mt}
\overline{M}(R)=\overline{R} \left[x(t_R) +\frac{\overline{R}}{6}
\left(\alpha(t_R)-\frac{\overline{R}}{24}\right) \right].
\ee
Recall that $\overline{M}(R)$ and $\overline{R}$ are defined in
Eq.(\ref{eq:bars}).

The analytical solution from Ref[\ref{Po}]
\be
\rhbar_1(t)=\frac{1}{\cosh^2(kt)}, \hskip 2cm k= \sqrt{\frac{2}{3}}
\label{eq:rho1}
\ee
is an extremely good approximation for $t \la 1$. The corresponding
$x(t)$ is
\be
x_1(t)= \frac{(kt) \tanh(kt)-t^2/[3\cosh^2(kt)]}
{(kt)\, \tanh(kt)+1}.
\ee
Eq.(\ref{eq:smain}) still has the exact solution $x=3/7$, which has the wrong
boundary conditions at $t=0$. However,
the solution with the correct boundary conditions at $t=0$ should approach
$x=3/7$ in the limit of large $t$, as indicated by our exact numerical
solutions.
To the second order
of a small parameter $A$, the perturbation around $x=3/7$ gives
\ba
\label{eq:x2}
x_2(t)&=&\frac{3}{7} +\frac{A}{t^{3/4}} \, \cos[\theta(t)]+\nonumber\\
&&\frac{7}{32}\, \frac{A^2}{t^{3/2}} \left\{-7+\frac{139}{54}\,
\cos[2\theta(t)]
-\frac{19}{54} \sqrt{47} \, \sin[2\theta(t)] \right\},
\ea
where
\be
\theta(t)= \frac{\sqrt{47}}{4} \ln \left(\frac{t}{3/7}\right)
+\tan^{-1}\left(\frac{1}{\sqrt{47}}\right)-\frac{\pi}{B}.
\ee
$A$ and $B$ are constants which are {\it independent} of $\rho(0)$. A good fit
of Eq.(\ref{eq:x2}) with the exact numerical solution gives
\be
A \simeq -0.18, \hskip 2cm B \simeq 3.54.
\ee
Eq.(\ref{eq:srho}) gives the corresponding $\rhbar(t)$ for $x_2(t)$:
\ba
\label{eq:rho2}
\rhbar_2(t)&= &\frac{1}{t^2} \left[ \frac{3}{7}+\frac{A}{4t^{3/4}} \, \left(
\cos[\theta(t)]-\sqrt{47} \, \sin[\theta(t)] \right) \right.\nonumber\\
& & \left.+\frac{7}{32} \frac{A^2}{t^{3/2}} \left( \frac{7}{2} -\frac{86}{9}
\cos[2\theta(t)]-\frac{10}{9} \sqrt{47} \,\sin[2\theta(t)] \right) \right].
\ea
Since Eq.(\ref{eq:rho1}) is valid for small $t$, while Eq.(\ref{eq:rho2}) is
valid for large $t$, we can construct an approximate solution by
combining the two solutions with appropriate weight functions, i.e.
\be
\label{eq:rhom}
\rhbar(t) = w_1(t) \, \rhbar_1(t)+w_2(t) \,\rhbar_2(t),
\ee
where $w_1(t)$ and $w_2(t)$ are weight functions. For example
\be
w_1(t)=\frac{1}{1+e^{t/n_1} \, t^{n_2}}, \hskip 2cm
W_2(t)= \frac{t^{n_2}}{e^{-t/{n_1}}+t^{n_2}}.
\ee
A good choice is
\be
n_1=20, \hskip 2cm n_2=8.
\ee

The cosmic balloon radius as a function of the central mass density
$\rho(0)$ can be computed by
using Eqs.(\ref{eq:rhom}) and (\ref{eq:Rq0}). We can write
\be
\label{eq:xm}
x(t)=w_1(t)\, x_1(t)+w_2(t) \,x_2(t).
\ee
The mass of the cosmic balloon is given by Eq.(\ref{eq:Mt}).

In Fig.1, we see that Eq.(\ref{eq:rhom}) is an extremely good approximation
of the exact mass density $\rhbar(t)$ (scaled to be independent of $\rho(0)$).
The radius $R$ and mass $M(R)$ obtained by using Eqs.(\ref{eq:rhom})
and (\ref{eq:xm}) are plotted in dashed lines in Figs. 2-4, they are quite
close to the exact solutions.

\section{6. Remarks}

The spiral behaviour in $M(R)$ versus $R$ of cosmic balloons (see Fig.4)
resembles that of fermion stars [\ref{LEE}]. This is not surprising since
cosmic balloons and soliton stars are similar objects.
$dM/d\rho(0)=0$ generally signifies the change from stability
to instability (or vice versa), and $dM/d\rho(0)>0$ indicates stability.
However, due to dynamical complications, the range of $M(R)$ and $R$
beyond the first mass maximum ($dM/d\rho(0)=0$)
on the curve in Fig. 4 does not correspond to stable configurations of cosmic
balloons, although the conditions for stability are satisfied each time the
curve bends upward (for increasing $\rho(0)$). It is possible that the
existence of the fixed point ($\overline{R}\simeq 0.468$, $\overline{M}(R)
\simeq 0.2275$, $2GM(R)/R \simeq 0.486$)
indicates that the cosmic balloons beyond the stability range oscillate
around a finite configuration with $2GM(R)/R \simeq 0.486$, instead of
collapsing
into black holes.

\centerline{\bf Acknowledgments}
I thank R. Caldwell and  V. Silveira helpful discussions.
This work was supported by the DOE and NASA under Grant NAGW-2381.

\newpage
\frenchspacing
\parindent=0pt

\centerline{{\bf References}}

\begin{enumerate}

\item\label{CB} B. Holdom, Univ of Toronto preprint \#UTPT-93-18 (1993).

\item\label{NB} B. Holdom, {\it Phys. Rev. D} {\bf 36} (1987) 1000; A.D. Dolgov
and O. Yu. Markin, {\it Sov. Phys. JETP} {\bf 71} (1990) 207; A.D. Dolgov and
O. Yu. Markin, {\it Prog. Theor. Phys.} {\bf 85} (1991) 1091; B. Holdom and
R.A. Malaney, CITA preprint CITA/93/22 (1993).

\item\label{Po} R. Ma\'nka, I. Bednarek, and D. Karczewska, Univ of Silesia
preprint U\'SL-TH-93-01.

\item\label{SIKIVIE} J. Ipser and P. Sikivie, {\it Phys. Rev. D} {\bf 30}
(1984) 712.

\item\label{SW} S. Weinberg, {\it Gravitation and Cosmology}, John Wiley and
Sons, 1972.

\item\label{LEE} T.D. Lee and Y. Pang, {\it Phys. Rev. D} {\bf 35} (1987)
(3678); T.D. Lee, {\it Phys. Rep.} {\bf 221} (1992) 251.

\end{enumerate}

\newpage
\nonfrenchspacing
\parindent=20pt

\centerline{{\bf Figure Captions}}

\vspace{0.2in}

Fig. 1. $\rhbar(t)$ and $\rhbar(t_R)=[3\alpha(t_R)+\beta^{-1}(t_R)]
/\left(4\sqrt{3q} \, t_R\right)$, with $t\equiv \sqrt{3q}\, \rbar$ and $q
\equiv 8\pi G \rho(0)r_0^2/3=0.9,2,10$
(dotted lines from right to left). The dashed line is the analytical
approximation of $\rhbar(t)$ from Eq.(\ref{eq:rhom}), it almost overlaps with
the exact solution (the solid line).

Fig. 2. $\overline{R}$ versus $q\equiv 8\pi G \rho(0)r_0^2/3$. The solid line
is the exact solution, while the dashed line is the result of analytical
approximation. The dotted line is $\overline{R}=0.468$.

Fig. 3. $\overline{M}(\overline{R})$ versus $q\equiv 8\pi G \rho(0)r_0^2/3$.
The solid line is the exact solution, while the dashed line is the result of
analytical approximation. The dotted line is $\overline{M}(R)=0.2275$.

Fig. 4. $M(R)$ versus $R$. The solid line
is the exact solution, while the dashed line is the result of analytical
approximation.

\end{document}